\documentclass[12pt,a4paper,final]{iopart}

\usepackage{iopams}   
\usepackage{graphicx, import}
\usepackage[breaklinks=true,colorlinks=true,linkcolor=blue,urlcolor=blue,citecolor=blue]{hyperref}

\newcommand{\be}{\begin{equation}}
\newcommand{\ee}{\end{equation}}
\newcommand{\bea}{\begin{eqnarray}}
\newcommand{\eea}{\end{eqnarray}}

\begin{document}

\title[Stochastic thermodynamics of rapidly driven systems]{Stochastic thermodynamics of rapidly driven systems}

\author{Gregory Bulnes Cuetara}
\address{Complex Systems and Statistical Mechanics, University of Luxembourg, L-1511 Luxembourg, Luxembourg}

\author{Andreas Engel}
\address{Fakult\"a V, Carl-von-Ossietzky Universit\"at Oldenburg, D - 26111 Oldenburg, Germany}

\author{Massimiliano Esposito}
\address{Complex Systems and Statistical Mechanics, University of Luxembourg, L-1511 Luxembourg, Luxembourg}


\begin{abstract}
We present the stochastic thermodynamics analysis of an open quantum system weakly coupled to multiple reservoirs and driven by a rapidly oscillating external field. The analysis is built on a modified stochastic master equation in the Floquet basis. Transition rates are shown to satisfy the local detailed balance involving the entropy flowing out of the reservoirs. The first and second law of thermodynamics are also identified at the trajectory level. Mechanical work is identified by means of initial and final projections on energy eigenstates of the system. We explicitly show that this two step measurement becomes unnecessary in the long time limit. A steady-state fluctuation theorem for the currents and rate of mechanical work is also established. This relation does not require the introduction of a time reversed external driving which is usually needed when considering systems subjected to time asymmetric external fields. This is understood as a consequence of the secular approximation applied in consistency with the large time scale separation between the fast driving oscillations and the slower relaxation dynamics induced by the environment. Our results are finally illustrated on a model describing a thermodynamic engine.
\end{abstract}

\tableofcontents


\section{Introduction}


The identification of thermodynamic quantities, such as heat, work and entropy, in open quantum systems driven by an external field is a central issue in quantum thermodynamics. Such systems are encountered in a variety of physical situations including the interaction with electromagnetic radiation \cite{Blumel_1991_PhysRevA}, driven tunneling \cite{Kohler_2005_PhysicsReports}, switching in multi-stable quantum systems \cite{Ketzmerick_2009_PhysRevE}, transport properties of driven quantum dots \cite{Wu_2010_PhysicalReviewB}, and non-equilibrium Bose-Einstein condensation \cite{Vorberg_2013_PhysRevLett}.

Up to now, a consistent picture of the thermodynamics of these systems has only been given within specific limits or regimes. In particular, most studies have been focused on slowly driven and weakly coupled open systems \cite{Esposito_2010_NewJournalofPhysics, Esposito_2010_EPL, Albash_2012_NewJournalofPhysics, Kosloff_2013_Entropy, Cuetara_2014_PhysicalReviewE}. Within this regime, the system dynamics is well described by a stochastic master equation in the basis of time dependent energies of the system. Entropy production, heat and work can then be identified at the single trajectory level, and the thermodynamic analysis of the system performed within the framework of stochastic thermodynamics (ST) \cite{Seifert_2012_ReportsonProgressinPhysics, Esposito_2012_PhysicalReviewE, VandenBroeck_2014_ArXive-prints}.

More recently, work has been done on the study of thermodynamic properties of open quantum systems driven by a fast and periodic external field, whether weakly coupled to a single heat reservoirs \cite{Wu_2010_PhysicalReviewB, Horowitz_2012_PhysicalReviewE, Zheng_2013_TheJournalofchemicalphysics, Langemeyer_2014_PhysicalReviewE, Gasparinetti_2014_arXivpreprintarXiv, Silaev_2014_PhysicalReviewE, Liu_2014_pre} or arranged in a specific implementation of a heat engine \cite{Gelbwaser-Klimovsky_2013_PhysicalReviewE, Szczygielski_2013_PhysicalReviewE} . In the present paper, we perform the general ST analysis of an open quantum system weakly coupled to multiple heat or chemical reservoirs, and driven by a fast and time periodic external force. Considering multiple reservoirs considerably widens the scope of possible applications such as, for example, the study of externally driven current or thermoelectric converters.

We perform the statistics of the energy and matter currents out of the reservoirs using the counting statistics formalism (see Ref. \cite{Esposito_2009_ReviewsofModernPhysics} for a review). Within this formalism, the currents are determined by making initial and final measurement of the energy and particle number in the reservoirs. In the weakly coupled and fast driving regime, these statistics is shown to be independent of quantum coherences in the Floquet basis. This directly results from the dynamical decoupling between populations and coherences in the Floquet basis in this regime, together with the fact that measurements of the reservoirs energy and particle number are independent of the system state.

The identification of the mechanical work further requires a double measurement of the initial and final energies of the system \cite{Tasaki_2000_arXivpreprintcond-mat, Kurchan_2000_arXivpreprintcond-mat, Talkner_2007_PhysicalReviewE, Campisi2011ReviewofModernPhysics}. Contrary to the current statistics, the mechanical work statistics depends on the evolution of coherences in the Floquet basis, which both influence and depend on the outcome of the system energy measurements. However, we show that the double projection in the system becomes unnecessary at steady-state for the identification of the rate of mechanical work, i.e. power. In this limit, the mechanical power statistics is exclusively determined by the diagonal elements of the modified density matrix in the Floquet basis, independently of quantum coherences. The first law then leads to a balance equation for the currents and the mechanical power. Furthermore, the steady-state mechanical power is shown to be given by the transfer rate of quanta, with energy given by the driving frequency ($\hbar = 1$), from the external driving to the reservoirs.

An important consequence of the dynamical decoupling between populations and coherences in the Floquet basis is that the trajectory entropy production associated to the stochastic dynamics in the Floquet basis satisfies a transient FT. The Shannon entropy in the Floquet basis is thus the relevant entropy within this scheme. It is remarkable that a FT relation for these systems can be derived without need to formally introduce a time reversed external driving \cite{Crooks_1998_JournalofStatisticalPhysics, Crooks_1999_PhysRevE, Seifert_2005_PhysicalReviewLetters}. This is a direct consequence of the assumption of large time scale separation between the fast driving oscillations and the slower relaxation time scale induced by the environment. Within this limit, a secular approximation over many driving oscillations can be consistently applied, resulting in a master equation with time-independent rates \cite{Blumel_1991_PhysRevA, Kohler_1997_PhysRevE, Kohler_1998_PhysRevE, Hone_2009_PhysicalReviewE, Langemeyer_2014_PhysicalReviewE}.

The connection between entropy production and the heat currents is provided by the local detailed balance (LDB) satisfied by the transition rates between Floquet states, which is here written in terms of the heat exchanged between the system and the reservoirs during the corresponding microscopic transition. The heat exchange includes multiples of the driving frequency which result from the presence of the non conservative external force due to the driving, and are identified as the dissipated mechanical work. We make use of the LDB condition in order to write the steady-state entropy production in terms of the currents and mechanical power. A steady-state FT for these quantities is also established by using the LDB, which is the steady-state version of the transient FT obtained in Ref. \cite{Cuetara_2014_PhysicalReviewE}. A steady-state FT for the mechanical power is recovered when considering a single heat reservoir \cite{Liu_2014_pre, Silaev_2014_PhysicalReviewE}.

This paper is organized as follows. In section \ref{Hamiltonian}, we first introduce the general Hamiltonian of a periodically driven open quantum system as well as the Floquet basis of the system and its associated quasi-energies. This section is mainly meant to fix notations. 

In section \ref{counting_statistics}, we perform the counting statistics of the currents of energy and matter through the system by using the counting statistics formalism. We derive the modified stochastic master equation \cite{Esposito_2009_ReviewsofModernPhysics} by using standard assumptions: weak coupling between system and environment, wide spacing between the quasi-energies and fast driving as compared to the relaxation dynamics \cite{Cohen-Tannoudji1988, Kohler_1998_PhysicalReviewE, Petruccione_Thetheoryofopenquantumsystems, Hone_2009_PhysicalReviewE, Langemeyer_2014_PhysicalReviewE}. This section extends former results \cite{Wu_2010_PhysicalReviewB, Horowitz_2012_PhysicalReviewE, Zheng_2013_TheJournalofchemicalphysics, Langemeyer_2014_PhysicalReviewE, Gasparinetti_2014_arXivpreprintarXiv} to an environment consisting of multiple reservoirs.

The ST analysis of the system starts in section \ref{stochastic_thermodynamics}. In the first part of this section we use the energy conservation law to construct the mechanical work statistics. The steady-state statistics is also discussed and the first law is introduced. In the second part, we show that the trajectory entropy production satisfies a transient FT and formally establish a FT for the currents and mechanical power.

We apply our results to the analysis of a thermodynamic engine in section \ref{models}. This engine consists of a two level system, weakly coupled to two particle reservoirs. For this system, the stochastic master equation is exposed and the large deviation function of the output power is numerically obtained and illustrated. We also investigate both average and fluctuations of the output power. Quite remarkably, the output power is subject to large fluctuations in the regime of maximum output power in this model.

Finally, a summary of the obtained results and possible perspectives are drawn in the concluding section \ref{conclusion}.


\section{Model Hamiltonian}\label{Hamiltonian}


We consider a periodically driven open quantum system modeled by a Hamiltonian of the form
\be \label{total_hamiltonian}
H (t)= H_{\rm S} (t) + H_{\rm R} + V,
\ee
where $H_{\rm S} (t) = H_{\rm S} (t+ T) $ denotes the Hamiltonian of the periodically driven system, $H_R$ stands for the environment Hamiltonian and $V$ describes the interaction between the system and the environment. 

According to Floquet theory, the dynamics associated to the time-periodic Hamiltonian $H_S (t)$ admit a complete set of solutions under the form $| \psi_s (t) \rangle = \mbox{e}^{-i \epsilon_s t} | s_t \rangle$, where $\epsilon_s $ are the so-called quasi-energies of the system while the Floquet states $|s_t \rangle$ have the same periodicity as the Hamiltonian, that is, $|s_{t+T} \rangle =|s_t \rangle$ \cite{Shirley_1965_PhysicalReview, ZelDovich_1967_SovietPhysicsJETP}. These Floquet states and quasi-energies satisfy the eigenvalue problem
\be
\left( H_{\rm S} (t) - i \partial_t \right) | s_t \rangle = \epsilon_s | s_t \rangle
\ee
which is obtained by inserting the quantum state $|\psi_s (t) \rangle$ into the Schr\" odinger equation associated to the system Hamiltonian $H_S (t)$. Floquet states can be Fourier expanded according to $| s_t \rangle = \sum_k e^{- i k \omega t} | s_k \rangle$ in terms of the driving frequency $\omega = 2 \pi / T$. The system quasi-energies are defined up to a multiple of the frequency $\omega$, and can thus be restricted to the first Brillouin zone, $\epsilon_s \in \left[ 0, \omega \right]$. 

We assume that the particle number operator in the system, denoted by $N_S$, commutes with the system Hamiltonian at all times, i.e. $\left[ N_S , H_S (t) \right] = 0$. As a result, the operators $ H_{\rm S} (t) - i \partial_t $ and $N_S$ can be simultaneously diagonalized and the Floquet states $|s \rangle$ may be chosen in such a way as to have a well defined particle number $n_s$.

The environment consists of a set of macroscopic reservoirs of energy and particles labelled by the index $\nu = 1 , \dots , N$. Its Hamiltonian is written as 
\be
 H_{\rm R}= \sum_{\nu }  H_{\nu} \quad \mbox{with} \quad H_\nu = \sum_{r} \epsilon_{r \nu} |  r\rangle_\nu \langle r |_\nu ,
\ee
where $ | r \rangle_\nu$ is a quantum state in the reservoir $\nu$ with energy $\epsilon_{r \nu} $ and particle number $n_{r\nu}$. The particle number operator in reservoir $\nu$ is then given by 
\be
N_{\nu}= \sum_{ r} n_{r \nu} |  r\rangle_\nu \langle  r |_\nu .
\ee 
Each reservoir $\nu$ is assumed to be initially at grand canonical equilibrium with inverse temperature $\beta_\nu = (k_{\rm B} T_\nu )^{-1} $ and chemical potential $\mu_\nu$
\be \label{equilibriumreservoir}
\rho^{eq }_\nu = \frac{\mbox{e}^{-\beta_\nu (H_\nu - \mu_\nu N_\nu)}}{Z_\nu} ,
\ee
where $Z_\nu = \mbox{Tr} \left\{ \mbox{e}^{-\beta_\nu (H_\nu - \mu_\nu N_\nu)} \right\}$ is the partition function of reservoir $\nu$.

The interaction between the system and its environment is written as
\be  \label{interactionV}
V = \sum_{\nu \kappa} S_{ \kappa } R^{\nu}_{\kappa} ,
\ee
where the sum runs over all the possible interaction terms and $ S_{\kappa}$ and $R_{\kappa}^{\nu}$ denote operators acting on the Hilbert space of the system and the reservoir $\nu$, respectively. 

The total particle number operator, $N = N_S + \sum_\nu N_\nu$, is assumed to commute with the total Hamiltonian (\ref{total_hamiltonian}), i.e. $\left[ N , H \right] = 0$, so that the total number of particles is conserved in the full system.


\section{Counting statistics of energy and matter currents} \label{counting_statistics}


At finite times, the statistical properties of the energy and matter currents are completely characterized by the generating function (GF)
\be \label{generating_function}
G(\xi_\nu , \lambda_\nu ,t) =  \langle \mbox{e}^{ - \sum_\nu \left( \xi_\nu \Delta \epsilon_\nu +\lambda_\nu \Delta n_\nu \right)} \rangle_t ,
\ee
the average $ \langle \cdot \rangle_t $ being taken with respect to the probability distribution $p (\Delta \epsilon_\nu , \Delta n_\nu ,t)$ of observing an amount of energy $\Delta \epsilon_\nu$ and particles $\Delta n_\nu$ flowing out of reservoir $\nu$ between time $0$ and time $t$.

The counting statistics formalism provides a general framework to calculate the GF (\ref{generating_function}) in open quantum systems. One introduces the modified Hamiltonian \cite{Esposito_2009_ReviewsofModernPhysics}
\be
H(\xi_\nu , \lambda_\nu , t) = \mbox{e}^{- \frac{i}{2} \sum_\nu \left( \xi_\nu H_\nu + \lambda_\nu N_\nu \right) } H (t) \mbox{e}^{ \frac{i}{2} \sum_\nu \left( \xi_\nu H_\nu + \lambda_\nu N_\nu \right) } ,
\ee

and the modified density matrix which satisfies the dynamical equation 
\be \label{generalizedmasterequation}
i \partial_t \rho (\xi_\nu , \lambda_\nu , t) =H(\xi_\nu , \lambda_\nu ,t) \rho (\xi_\nu , \lambda_\nu , t) - \rho (\xi_\nu , \lambda_\nu , t) H(-\xi_\nu , -\lambda_\nu ,t).
\ee
The current GF can then be written as the trace of the modified density matrix, $G(\xi_\nu , \lambda_\nu , t) = \mbox{Tr} \left\{  \rho ( i \xi_\nu , i \lambda_\nu , t) \right\}$. 

We now proceed by making the standard assumptions leading to a stochastic master equation for the diagonal elements in the Floquet basis of the reduced density matrix of the system 
\be
\rho_S (\xi_\nu , \lambda_\nu , t) \equiv \mbox{Tr}_{R} \left\{ \rho (\xi_\nu , \lambda_\nu , t)  \right\},
\ee 
the trace $\mbox{Tr}_R \left\{ \cdot \right\}$ being taken over the reservoirs degrees of freedom. A similar approach has recently been used in order to study the thermodynamics of rapidly driven quantum systems connected to a single heat reservoir \cite{Langemeyer_2014_PhysicalReviewE, Gasparinetti_2014_arXivpreprintarXiv}.

The whole system is assumed to be initially in the factorized state
\be
\rho (\xi_\nu , \lambda_{\nu} 0) = \rho_S (0) \prod_{\nu} \otimes \rho_{\nu}^{eq} ,
\ee
where $\rho_S (0)$ denotes the initial reduced density matrix of the system and the $\rho_{\nu}^{eq}$ are defined in (\ref{equilibriumreservoir}). We further make the following assumptions \cite{Blumel_1991_PhysRevA, Kohler_1997_PhysRevE, Kohler_1998_PhysRevE, Cohen-Tannoudji1988, Kohler_1998_PhysicalReviewE, Petruccione_Thetheoryofopenquantumsystems, Hone_2009_PhysicalReviewE, Langemeyer_2014_PhysicalReviewE}:
\begin{enumerate}
\item
The environment is composed of reservoirs which are weakly coupled to the quantum system and large enough to remain unaffected by the quantum system. Their correlation time $\tau_C$ is then assumed to be much shorter than the typical relaxation time scale of the system $\tau_R$.
\item
The free oscillations at the driving frequency, $\omega$, and at the Bohr frequencies of the Floquet basis, $\omega_{ss'} = \epsilon_s - \epsilon_{s'}$, are much faster than the relaxation process induced by the reservoirs over time scale $\tau_R$. We note that since quasi-energies are restricted to the first Brillouin zone, $\epsilon_s - \epsilon_{s'} \leq \omega$, a fast driving frequency is necessary though not sufficient in order to have a sparse Floquet spectrum. We further note that the absence of degeneracies in the energies of the undriven system does not necessarily imply the absence of degeneracies in the Floquet spectrum. A careful analysis of the Floquet spectrum is thus necessary in order to check the validity of the present assumption. We refer the reader to Refs. \cite{Hone_2009_PhysicalReviewE, Langemeyer_2014_PhysicalReviewE} for a more detailed discussion on this account.
\end{enumerate}

Under these assumptions, one can take the Born-Markov approximation and apply the rotating wave approximation (RWA) \cite{Cohen-Tannoudji1988, Petruccione_Thetheoryofopenquantumsystems} by averaging the system dynamics over a time scale $\Delta  t$ which is intermediate between 
\be
\tau_C \ll  \Delta t \ll \tau_R.
\ee
As a result of this procedure, the dynamics of the populations and coherences in the Floquet basis are decouple. The GF of the currents is then completely determined by the diagonal elements of the system reduced density matrix
\be \label{GFpop}
G(\xi_\nu , \lambda_\nu ,t) = \sum_{s} \rho_{ss} (i\xi_\nu , i \lambda_\nu ,t),
\ee
where $\rho_{ss'} (\xi_\nu , \lambda_\nu ,t) = \langle s | \rho_S (\xi_\nu , \lambda_\nu ,t) | s' \rangle$.

We first give the modified stochastic master equation that rule the evolution of populations $g_s (\xi_\nu , \lambda_\nu ,t) \equiv \rho_{ss} ( i \xi_\nu , i \lambda_\nu ,t)$. In the following, functions defined on the set of quasi-energy states ${\bf f} : s \rightarrow f_s$ are arranged into vectors with components $\left[ {\bf f} \right]_s = f_s$. For brevity, the sum of their components are written as $f \equiv \sum_s f_s = {\bf 1} \cdot {\bf f}$ where ${\bf 1} \equiv (1 ,1 , \dots , 1)$ and $\cdot$ denotes a matrix product.

With these notations, populations in the Floquet basis follow the set of dynamical equations
\be \label{modmasteq}
\dot {\bf g}  ( \xi_\nu ,   \lambda_\nu , t ) = {\bf \Gamma} ( \xi_\nu ,   \lambda_\nu )  \cdot  {\bf g}  (\xi_\nu ,  \lambda_\nu , t ) 
\ee
where the matrix elements $ {\bf \Gamma} ( \xi_\nu , \lambda_\nu )$ containing the counting parameters can be written as
\be\label{ratematrix}
\left[ {\bf \Gamma}  (\xi_\nu ,   \lambda_\nu ) \right]_{ss'} = \sum_{ \nu , l} \left( \Gamma^{\nu l}_{ss'} \mbox{e}^{- \xi_\nu (\epsilon_{s'} - \epsilon_{s} + l \omega)} \mbox{e}^{- \lambda_\nu (n_{s'} - n_{s})}  - \delta_{ss'} \sum_{\tilde s} \Gamma^{\nu l}_{\tilde s s'} \right).
\ee

The transition rates appearing in (\ref{ratematrix}) are given by
\be \label{transitionrates}
\Gamma_{ss'}^{\nu l} = \sum_{\kappa  \kappa'}  \gamma_{\kappa \kappa' | ss'}^{l}   \alpha_{\kappa' \kappa}^{\nu} (\epsilon_{s'} - \epsilon_s + l \omega).
\ee

In this last expression, the amplitudes
\bea \label{amplitudes}
 \gamma_{\kappa \kappa' | ss'}^{l} & = & \sum_{k_1 k _2} \langle s_{k_1} | S_{\kappa} | s'_{k_1 + l} \rangle \langle s'_{k_2 + l} | S_{\kappa'} | s_{k_2} \rangle \\
& = & \left(  \int_{0}^{T} \frac{d t}{T} \mbox{e}^{i \omega l t} \langle s_t | S_{\kappa} | s'_t \rangle \right) \left( \int_{0}^{T} \frac{d t}{T} \mbox{e}^{ - i \omega l t} \langle s'_t | S_{\kappa'} | s_t \rangle    \right) \label{amplitudes2}
\eea
characterize the time scale of the corresponding transitions and depend on the number of quanta $l$ transferred from the driving protocol to the reservoirs. An important feature of these amplitudes is that they do not necessarily vanish for $s=s'$ leading to so-called pseudo-transitions between different modes of the same Floquet state \cite{Langemeyer_2014_PhysicalReviewE}. These pseudo-transitions directly contribute to the statistics of the currents which is manifest by the presence of terms of the form $\mbox{e}^{ \xi_\nu l \omega}$ along the diagonal of the transition rate matrix (\ref{ratematrix}).

Furthermore, relation (\ref{amplitudes2}) emphasizes the fact that the allowed number of quanta exchanged with the mechanical driving during stochastic transitions is determined by the spectral properties of the matrix elements $\langle s_t | S_{\kappa} | s'_t \rangle$. 

The reservoir correlation functions, on the other hand, are given by
\be \label{correlationfunctions}
\alpha^{\nu}_{\kappa \kappa'} (x) = \int_{-\infty}^{\infty} d\tau \mbox{e}^{i x \tau}  \mbox{Tr}_\nu \left\{ R^{\nu}_{\kappa} (\tau) R^{\nu}_{\kappa'} \rho_\nu \right\} ,
\ee
with $\rho_\nu$ denoting the grand canonical equilibrium density matrix (\ref{equilibriumreservoir}) in reservoir $\nu$ and the trace $\mbox{Tr}_\nu \left\{  \cdot \right\}$ being taken over its Hilbert space. These equilibrium correlation functions encapsulate the thermodynamic properties of the reservoir and satisfy the Kubo-Martin-Schwinger (KMS) condition
\be \label{KMS}
\alpha^{\nu}_{\kappa \kappa'} (x) = \alpha^{\nu}_{\kappa' \kappa } (-x) \mbox{e}^{\beta_\nu (x -\mu_\nu \Delta n^{\nu}_{\kappa} )} ,
\ee
where $\Delta n^{\nu}_\kappa$ denotes the particle number change in reservoir $\nu$ induced by the operator $R_{\kappa}^{\nu}$, that is, assuming that $\langle r | R_{\kappa}^{\nu} | r' \rangle \propto \delta (n_{r} - n_{r'} - \Delta n^{\nu}_\kappa) $.

The coherences in the Floquet basis, $\rho_{ss'}(\xi_\nu , \lambda_\nu ,t)$ with $ s\neq s'$, are also shown to follow the dynamics
\be \label{coherencesevol}
\dot \rho_{ss'}(\xi_\nu , \lambda_\nu ,t) = (- \Upsilon_{ss'} (\xi_\nu) - i \Theta_{ss'}) \rho_{ss'}(\xi_\nu , \lambda_\nu ,t) ,
\ee
with damping rates given by
\bea \nonumber \fl
\Upsilon_{ss'} (\xi_\nu) & = & - \sum_{l , k_1, k_2} \sum_{\kappa \kappa'} \langle s_{k_1} | S_\kappa | s_{k_1 + l} \rangle \langle s'_{k_2 +l} | S_{\kappa '} | s'_{k_2 } \rangle  \alpha_{\kappa' \kappa}^{\nu} (l \omega) \mbox{e}^{i \xi_\nu l \omega } \\ \fl
& & + \frac{1}{2}\sum_{l  \tilde s} \sum_{\kappa \kappa'} \left( \gamma_{\kappa \kappa' | s \tilde s}^{l} \alpha^{\nu}_{\kappa \kappa'} (\epsilon_s - \epsilon_{\tilde s} - l \omega) + \gamma_{\kappa \kappa' | s' \tilde s}^{l} \alpha^{\nu}_{\kappa \kappa '} (\epsilon_{s'} - \epsilon_{\tilde s} - l \omega) \right) \label{dampingrates}
\eea
and frequencies by
\bea  \nonumber \fl
\Theta_{ss'} & = & \epsilon_s - \epsilon_{s'} \\ \fl
&  + & \frac{1}{2 \pi}\sum_{l \tilde s} \sum_{\kappa \kappa'} \mbox{p.v.} \int_{-\infty}^{\infty} dx \left( \gamma^{l}_{\kappa \kappa'  | s \tilde s} \frac{\alpha^{\nu}_{\kappa \kappa'} (x)}{\epsilon_{s} - \epsilon_{\tilde s} - l \omega - x} -  \gamma^{l}_{\kappa \kappa'  | s' \tilde s} \frac{\alpha^{\nu}_{\kappa \kappa'} (x)}{\epsilon_{s'} - \epsilon_{\tilde s} - l \omega - x} \right)
\eea
where $\mbox{p.v.}$ denotes the Cauchy principal value.

The coherences thus evolve independently of each other and undergo exponentially damped oscillations. Quite remarkably the damping rates (\ref{dampingrates}) depend on the energy counting fields $\xi_\nu$ contrary to the autonomous situation.

The amount of energy and matter exchanged between the system and reservoirs during a microscopic transition is apparent in the expression of the modified rates (\ref{ratematrix}). The different transitions between states $s$ and $s'$ are distinguished by their indices $\nu$ and $l$. Such transitions involve an energy and particle exchange between the system and reservoir $\nu$ respectively given by $ \epsilon_{s} - \epsilon_{s'} - l \omega$ and $  n_{s} - n_{s'}$.

The summation over integer multiples of the driving frequency $\omega$ in the rate matrix elements (\ref{ratematrix}) is characteristic of the presence of the external periodic force. As we will later see, the non conservative contributions $l \omega$ to the energy flow are identified as the mechanical work dissipated into the reservoirs at steady-state. The statistics of the mechanical work is then obtained in the long time limit by only counting these terms.

At this point, we note that the usual stochastic master equation in the Floquet basis \cite{Blumel_1991_PhysRevA, Kohler_1997_PhysRevE, Kohler_1998_PhysRevE, Hone_2009_PhysicalReviewE, Langemeyer_2014_PhysicalReviewE} is simply obtained by setting the counting fields to zero in equations (\ref{modmasteq}) and (\ref{coherencesevol}), i.e. $\rho_{ss'} (t) =\left. \rho_{ss'} (\xi_\nu , \lambda_\nu ,t)\right|_{\xi_\nu = \lambda_\nu = 0}$.  Such equation is the analog of the stochastic master equation derived in within the Born-Markov and secular approximations for autonomous systems. The main difference in the driven case is the appearance of quasi-energies which replace the system energies of the autonomous case, and the appearance of integer multiples of the driving frequency in the amounts of energy exchanged between the system and the reservoirs. In the absence of external driving, the quasi-energies $\epsilon_s$ become the actual energies of the system and the summation over $l$ disappears, leading to the usual master equation for autonomous open systems.

An essential task in ST is the identification of the microscopic processes related through time reversal. Such processes involve opposite amounts of energy and particle exchanges with the environment as well as inverted initial and final states. From the above discussion, these pairs of processes have transition rates  given by $ \Gamma^{\nu l}_{ss'} $ and $ \Gamma^{\nu -l}_{s's} $. 

By virtue of the symmetry relation $ \gamma_{\kappa \kappa' | ss'}^{l}  = \gamma_{\kappa' \kappa | s's}^{-l}  $ and KMS condition (\ref{KMS}), these pairs of transition rates satisfy the LDB condition
\be \label{detailedbalance}
\ln \frac{\Gamma_{ss'}^{\nu l}}{  \Gamma_{s's}^{\nu \, -l} } =-\beta_\nu ( \epsilon_{s} - \epsilon_{s'} - l \omega - \mu_\nu (n_{s} - n_{s'})),
\ee
where the right-hand side is the entropy flowing from reservoir $\nu$ during the transition.

The presence of $l \omega$ terms in the energetics of (\ref{detailedbalance}) shows that the mechanical driving can enhance or decrease the statistical frequency of particular transitions. For example, by providing an extra amount of energy through the exchange of quanta at the driving frequency, the mechanical driving effectively lowers the energy cost of a particular transition thus increasing its probability rate. This observation will prove useful in the study of the thermodynamic engine considered in section \ref{models}.

Finally, we note that the quantity $- \mu_\nu (n_{s} - n_{s'})$ is the chemical work performed by the system to bring $ n_{s} - n_{s'} $ particles into reservoir $\nu$ against the chemical potential $\mu_\nu$.

The fundamental relation (\ref{detailedbalance}) plays a key role in writing the entropy production as the sum of the system entropy change and the entropy flow from the environment \cite{Esposito_2007_PhysRevE, Esposito_2012_PhysicalReviewE, Seifert_2012_ReportsonProgressinPhysics}. In addition, it also leads to a steady-state FT for the mechanical work and the currents out of the reservoirs as we show in section \ref{entropy_balance}.

For systems maintained in a non-equilibrium steady-state by boundary constraints, such as temperature and chemical potential differences between the reservoirs, the cumulant generating function (CGF)
\be \label{CGF}
\mathcal{G} (\xi_\nu , \lambda_\nu)  \equiv   \lim_{t \rightarrow \infty} \frac{1}{t} \ln G(\xi_\nu , \lambda_\nu ,t) 
\ee
is a measure of the current fluctuations at steady-state. In particular, all the moments and correlations between the currents can be obtained by successive derivation of the CGF with respect to its counting parameters $\xi_\nu$ and $\lambda_\nu$ at zero values.

A related object is the large deviation function (LDF) of the currents
\be \label{LDF}
\mathcal{I} (j_{\nu}^{\epsilon} , j_{\nu}^{n}) \equiv - \lim_{t \rightarrow \infty} \frac{1}{t} \ln p (\Delta \epsilon_{\nu} , \Delta n_\nu ,t) ,
\ee
where the currents are defined as 
\be \label{currentstochvar}
j_{\nu}^{\epsilon} = \frac{\Delta \epsilon_\nu}{t} \quad \mbox{and} \quad j_{\nu}^{n} = \frac{\Delta n_\nu}{t}.
\ee
The CGF (\ref{CGF}) and LDF (\ref{LDF}) are related through the Legendre-Fenchel transformation as stated by the G\"artner-Ellis theorem \cite{Touchette_2009_PhysicsReports}.

Using the formal solution of equation (\ref{modmasteq}), the current GF can be written as 
\bea
G (\xi_\nu , \lambda_\nu ,t ) & = &  {\bf 1} \cdot \mbox{e}^{{\bf \Gamma}  (\xi_\nu ,   \lambda_\nu ) t} \cdot {\bf p }_0 
\eea
where ${\bf p }_0$ denotes the initial occupation probability of the system. This also shows that the CGF (\ref{CGF}) is given by the dominant eigenvalue of the rate matrix ${\bf \Gamma}  (\xi_\nu ,   \lambda_\nu )$ \cite{Esposito_2009_ReviewsofModernPhysics}. Besides, the average values of the energy and matter currents, obtained as the first derivatives of the CGF, are then given by
\bea \label{energycurrent}
&& J_{\nu}^{\epsilon} = -\partial_{\xi_\nu}\mathcal{G} (0,0) =  \sum_{\nu l }  \sum_{ss'} \left( \epsilon_s - \epsilon_{s'} - l \omega \right) \Gamma_{ ss'}^{\nu l}  p^{st}_{s'}  \\ \label{particlecurrent}
&& J_{\nu}^{n} = -\partial_{\lambda_\nu}\mathcal{G} (0,0)   =   \sum_{\nu l }  \sum_{ss'} (n_s - n_{s'})   \Gamma_{ ss'}^{\nu  l} p^{st}_{s'},
\eea
in terms of the steady-state probabilities ${\bf p^{st}} = \lim_{t \rightarrow \infty} {\bf p} (t)$.

Finally, let us mention some interesting differences between the fast driving limit considered here and the slow driving limit. In this latter case, populations of the density matrix are known to satisfy a stochastic master equation in the time dependent energy eigenbasis of the system. The presence of the external field is then manifest by the time dependent system energies appearing in the tunnelling rates. These rates are known to satisfy a LDB condition, which depends on the time dependent parameters of the system. In the rapidly driven systems considered here, populations and coherences of the density matrix are dynamically decoupled in the Floquet basis and the stochastic master equation is now time independent in this basis. In this regime, the external driving results in the presence of non conservative terms in the LDB, which are expressed under the form of integer multiples of the driving frequency.


\section{Stochastic Thermodynamics}\label{stochastic_thermodynamics}


The whole framework of ST relies on the identification of the first and second laws at the microscopic level. The first law requires the discrimination between the mechanical and thermal contributions to the energy balance of the considered physical system. The microscopic version of the second law is expressed as a transient FT for the trajectory entropy production.

In the following subsection we identify the mechanical work by using initial and final measurements of the system energy. Its statistics is derived and particular emphasis is put on the steady-state fluctuations of mechanical power. Within this limit, the initial and final measurements of energy are shown to be irrelevant and the mechanical power can then be interpreted as the transfer rate of quanta to the system at the driving frequency.

The second law and FTs are discussed in  subsection \ref{entropy_balance}. Since populations and coherences in the Floquet basis are decoupled in the regime considered here, the trajectory entropy production of the stochastic process ruled by (\ref{modmasteq}) and (\ref{ratematrix}) satisfies a transient FT  \cite{Seifert_2005_PhysicalReviewLetters, Esposito_2010_PhysRevLett, Seifert_2012_ReportsonProgressinPhysics, VandenBroeck_2014_ArXive-prints}. Quite remarkably, this is true despite the quantum coherences in the Floquet basis introduced by the initial measurement of the system energy. We further consider the long time limit and formally establish a steady-state FT for the currents and mechanical power \cite{Cuetara_2014_PhysicalReviewE}.


\subsection{Energy balance and work statistics}\label{energy_balance}


In the weak coupling limit, the mechanical work performed by the external driving is given by the changes in system and reservoir energies between initial and final times, respectively chosen as time $0$ and time $t$. Measuring the energy change in the system requires projective measurements of its initial and final energies. The necessity to project the system at initial and final times in order to perform the energetic analysis stems from the fact that Floquet states are not eigenstates of the time dependent system Hamiltonian $H_S (t)$.

In the following, we denote by $| e_t \rangle$ the instantaneous eigenstate of the system Hamiltonian $H_s (t)$ with eigenvalue $e_t$, that is $H_S (t) | e_t \rangle= e_t | e_t \rangle$. The system is assumed to undergo ideal measurements of its energy at initial and final times yielding the values $e_0$ and $e_t$ respectively. The reduced density matrix of the system is thus given by $\rho_S (0 ; e_0) \equiv | e_0 \rangle \langle e_0 | $ in case the initial measurement of the system energy yields $e_0$, while this happens with probability $p_{e_0} =  \langle e_0 | \rho_S (0) | e_0 \rangle$.

The mechanical work is then given by the changes in the system and environment energies, i.e. $w = \Delta e_S - \sum_\nu \Delta \epsilon_\nu$, with $\Delta e_S = e_t - e_0$ and $\Delta \epsilon_\nu$ denoting the change of energy in reservoir $\nu$ between times $0$ and $t$. By following the general approach exposed in Ref. \cite{Esposito_2009_ReviewsofModernPhysics}, we obtain the generating function of the work as
\bea 
G_w ( \alpha , t ) & = & \langle \mbox{e}^{- \alpha w} \rangle_t  \\
& = &  \sum_{e_0 e_t} \mbox{e}^{- \alpha \Delta e_S} \sum_{ss'} \langle e_t | s' \rangle   \rho_{s's} (i\alpha, t; e_0) p_{e_0}  \langle s| e_t \rangle . \label{transientworkGF}
\eea
The average in the first line is taken with respect to the work distribution $p (w , t)$ of observing an amount of work $w$ performed by the external driving from time $0$ to $t$. On the second line, the modified density matrix elements $ \rho_{s's} (\alpha, t; e_0)$ are obtained from those of the modified rate matrix of the currents introduced in the previous section as $  \rho_{s's} (\alpha, t; e_0) = \left. \rho_{s's} (\xi_\nu , \lambda_\nu  , t) \right|_{\xi_\nu  = - \alpha,  \lambda_\nu = 0}  $. Note that the initial condition used in order to solve the dynamical equations (\ref{modmasteq}) and (\ref{coherencesevol}) is now to be taken as
\be \label{initialconditionworkGF}
  \rho_{s's} ( 0 ; e_0)  = \langle s' | \rho_S (0 ; e_0) | s\rangle 
\ee
due to the initial measurement of the system energy.

We further note that this initial measurement of the system energy also affects the current statistics at finite times. Indeed, if an initial measurement of the system energy is performed, one must consider the initial condition $g_{s} (\xi_\nu , \lambda_\nu ,0) = \sum_{e_0}  \rho_{ss} ( 0 ; e_0) $ when solving the dynamical equations (\ref{modmasteq}). However, though coherences in the Floquet basis of the system density matrix $\rho_S (0)$ may affect the initial weight of the populations after the measurement as taken place, the GF (\ref{generating_function}) is independent of the subsequent evolution of coherences in the Floquet basis induced by this measurement.

To the contrary, the mechanical work GF (\ref{transientworkGF}) does depend on the coherences in the Floquet basis induced by the initial measurement. This is mainly due to the fact that the operator which is counted in order to perform the work statistics, $H_S (t) + \sum_\nu H_\nu$, does not necessarily commute with the initial density matrix of the system $ \rho_S (0)$ before the first measurement has been performed \cite{Esposito_2009_ReviewsofModernPhysics}.

Nevertheless, the steady-state CGF of the mechanical work 
\be \label{workCGF}
\mathcal{G}_w (\alpha) = \lim_{t \rightarrow \infty} \frac{1}{t} \ln G_w (\alpha , t)
\ee
only depends on the populations of the modified density matrix $\rho_{ss} (\alpha, t; e_0) $ since coherences vanish at steady-state, i.e. $\lim_{t\rightarrow \infty} \rho_{ss'} (t) = 0$ for $s\neq s'$ (see the appendix for details). Provided the energy in the system remains finite in the long time limit, the mechanical work CGF (\ref{workCGF}) is then obtained as the dominant eigenvalue of the rate matrix  
\be \label{wordrm}
\left[ {\bf \Gamma}  (\alpha ) \right]_{ss'} = \sum_{ \nu , l}  \Gamma^{\nu l}_{ss'} \mbox{e}^{\alpha l \omega}- \delta_{ss'} \sum_{\nu , l , \tilde{s}} \Gamma^{\nu l}_{\tilde{s} s'},
\ee
obtained by making the substitutions $\xi_\nu \rightarrow \alpha$ and $\lambda \rightarrow 0$ in the rate matrix (\ref{ratematrix}) and noting that terms of the form $\mbox{e}^{\alpha (\epsilon_{s'} - \epsilon_s)}$ do not contribute to its eigenvalue.

At the trajectory level, we see that the stochastic variable associated to the mechanical power $\dot w$ is given by the transfer rate of quanta from the external driving to the system multiplied by the driving frequency, i.e. $\dot w \sim \omega \Delta l / t$ for $t \rightarrow \infty$ and where $\Delta l$ denotes the number of quanta transfered from the driving during a given realization of the dynamics. At steady-state, this mechanical power is entirely dissipated into the reservoirs.

The above discussion also shows that the mechanical power CGF (\ref{workCGF}) can be obtained from the current CGF (\ref{currentCGF}) by the following substitution
\be
\mathcal{G}_w (\alpha) = \mathcal{G} (\alpha , 0).
\ee
This relation emphasizes the fact that, at steady state, the mechanical power is equal to the sum of incoming energy currents from the reservoirs, that is, $\dot w = \sum_\nu j_{\nu}^\epsilon$.

We are now in position to write down the first law of thermodynamics at steady-state, relating the heat currents to the mechanical and chemical powers. By introducing the heat flows $\dot q_\nu =  j^{\epsilon}_\nu -  \mu_\nu j^{n}_\nu$ in terms of the currents (\ref{currentstochvar}), as well as the chemical power $\dot w_c =\sum_\nu \mu_\nu j^{n}_\nu$, the first law of thermodynamics reads
\be \label{firstlawtraj}
\sum_\nu  \dot q_\nu + \dot w_c + \dot w = 0
\ee
at steady-state, that is, for $t \rightarrow \infty$.

We note that the average rate of mechanical work is obtained from (\ref{workCGF}) as
\be \label{meanwork}
\dot W =   -\partial_{\alpha}\mathcal{G}_w (0)  = \omega   \sum_{\nu l }  \sum_{ss'} l \,  \Gamma_{ ss'}^{\nu l}  p^{st}_{s'},
\ee
which is the steady-state current of quanta with frequency $\omega$ injected into the system. A direct inspection of this relations together with (\ref{energycurrent}) and (\ref{particlecurrent}) shows that $\dot W = \sum_\nu J_{\nu}^{\epsilon}$. This relation can also be used in order to write the first law at the average level, in consistency with (\ref{firstlawtraj}),
\be
\sum_\nu  \dot Q_\nu + \dot W_c + \dot W = 0
\ee
where the average heat flow out of reservoir $\nu$ is given by $ \dot Q_\nu = J_\nu - \mu_\nu J_\nu$, and the rate of chemical work provided to the system by particles flowing out of the reservoirs by $ \dot W_c = \sum_\nu \mu_\nu J_\nu$.


\subsection{Entropy balance and fluctuation theorem} \label{entropy_balance}


The populations of the system in the Floquet basis satisfy a closed stochastic master equation \cite{Blumel_1991_PhysRevA, Kohler_1997_PhysRevE, Kohler_1998_PhysRevE, Hone_2009_PhysicalReviewE, Langemeyer_2014_PhysicalReviewE} as can be seen by setting the counting fields to zero in Eqs. (\ref{modmasteq}) and (\ref{transitionrates}). 
Since transition rates satisfy the LDB (\ref{detailedbalance}), the trajectory entropy production associated to this stochastic process can be decomposed into \cite{Esposito_2007_PhysRevE, Esposito_2010_PhysRevLett}
\be
\Delta_i s = \Delta s - \Delta_e s
\ee
where $\Delta s$ denotes the change in the system entropy and $\Delta_e s = \sum_\nu \beta_\nu \Delta q_\nu$ is the entropy flow from the environment. The probability distribution of the entropy production in a system ruled by a stochastic master equation is known to satisfy a fundamental FT at finite times  \cite{Seifert_2005_PhysicalReviewLetters, Esposito_2010_PhysRevLett, Seifert_2012_ReportsonProgressinPhysics, VandenBroeck_2014_ArXive-prints}.
The fact that this result applies in our case thus simply follows from the dynamical decoupling between populations and coherences in the Floquet basis when the driving frequency is sufficiently high.


Let us however emphasize a striking difference between systems driven by rapidly oscillating fields as considered here and those driven by a slow and/or non-periodic external driving $\lambda (t)$. In the latter case, definiting the trajectory entropy production requires the introduction of backward trajectories which are assumed to be ruled by the backward dynamics defined along a time reversed external driving $\lambda (\tau -t)$, where $\tau$ denotes the time length of the considered trajectory. In the present case however, such inversion of the external protocol is not needed since the generator of the stochastic process is effectively time independent. This is a direct consequence of the time averaging over many driving periods (RWA) and the large time scale separation between the fast driving oscillations and the slower relaxation process induced by the reservoirs.

At steady-state, the entropy change in the system becomes negligible as compared to the entropy flow from the environment. As a result, the rate of entropy production becomes equal to the rate of entropy flow from the environment in this limit. The FT for the entropy production then leads to a steady-state FT for the currents, independently of the initial condition in the system.

Such steady-state FT for the currents is now proven by establishing a fluctuation symmetry for the current CGF (\ref{CGF}). As a first step, we note that the rate matrix (\ref{transitionrates}) satisfies
\be 
{\bf \Gamma}  (\xi_\nu ,   \lambda_\nu ) = {\bf \Gamma}  (\beta_\nu - \xi_\nu , - \beta_\nu \mu_\nu -  \lambda_\nu )^{\top},
\ee
by virtue of the LDB condition (\ref{detailedbalance}) and where $^{\top}$ denotes a matrix transposition. Since the CGF is obtained as the dominant eigenvalue of the modified rate matrix, this last relation leads in turn to the aforementioned fluctuation symmetry
\bea \label{currentCGF}
\mathcal{G} (\xi_\nu , \lambda_\nu) =  \mathcal{G} ( \beta _ \nu - \xi_\nu , - \beta_\nu \mu_\nu - \lambda_\nu ).
\eea

This FT can be equivalently restated in terms of the large deviation function of the currents as \cite{Esposito_2009_ReviewsofModernPhysics}
\be \label{FT2}
\mathcal{I} (j_{\nu}^{\epsilon} , j_{\nu}^{n}) - \mathcal{I} (-j_{\nu}^{\epsilon} ,- j_{\nu}^{n})= \sum_\nu \beta_\nu \left( j_{\nu}^{\epsilon} - \mu_{\nu} j_{\nu}^{n} \right)
\ee
where the stochastic variables $j_{\nu}^{\epsilon} $ and $j_{\nu}^{n} $ stand for the steady-state currents of energy and matter, respectively, flowing out of reservoir $\nu$.

Alternatively, the symmetry relation (\ref{currentCGF}) leads to a FT for the currents and the mechanical power making explicit reference to the thermodynamic affinities applied to the system. By using the fact that the rate of power is equal to the sum of energy currents incoming from the reservoirs at steady-state, we note that a CGF of the work and currents can be obtained by making the following substitution in the counting fields
\be
\tilde \mathcal{ G} (\alpha , \chi_\nu , \eta_\nu ) \equiv \left. \mathcal{G} (-\alpha + \chi_\nu , \eta_\nu) \right|_{\chi_1 = \eta_1 = 0}.
\ee
In this last relation, the counting field $\alpha$ accounts for the mechanical power fluctuations. The symmetry relation (\ref{currentCGF}) then leads to the steady-state FT
\be \label{FTcurrentwork}
\tilde \mathcal{ G} (\alpha ,  \chi_\nu , \eta_\nu) \equiv \tilde \mathcal{  G} (\beta_1 - \alpha , A^{\epsilon}_{\nu} - \chi_\nu , A^{n}_{\nu} - \eta_\nu),
\ee
for the mechanical work and current fluctuations, and in terms of the thermodynamic forces driving the currents
\be \label{affinities}
A^{\epsilon}_\nu = \beta_1 - \beta_\nu \quad \mbox{and} \quad  A^{n}_\nu = \beta_\nu \mu_\nu - \beta_1 \mu_1.
\ee

Again, this FT is equivalent to
\be \label{LDFFT}
\mathcal{I} ( \dot w, j_{\nu}^{\epsilon} , j_{\nu}^{n}) - \mathcal{I} (-\dot w , -j_{\nu}^{\epsilon} ,- j_{\nu}^{n})=\dot w + \sum_\nu  \left(A_{\nu}^{\epsilon} j_{\nu}^{\epsilon} + A_{\nu}^{n} j_{\nu}^{n} \right)
\ee
in terms of the large deviation function of the mechanical power and currents
\be
\mathcal{I} (\dot w ,j_{\nu}^{\epsilon} , j_{\nu}^{n}) = \sup_{\alpha , \xi_\nu, \lambda_\nu}
\left\{ \dot w \alpha + \sum_{\nu = 2}^{N} \left( j_{\nu}^{\epsilon} \xi_\nu +  j_{\nu}^{n} \lambda_\nu  \right) - \tilde \mathcal{ G} (\alpha , \xi_\nu , \lambda_\nu)   \right\}.
\ee
This FT is the steady-state version of the finite-time FT for the work and currents obtained in Ref. \cite{Cuetara_2014_PhysicalReviewE}. The presence of a FT for the current fluctuations is known to have important consequences on the response properties of the system \cite{Andrieux_2004_TheJournalofchemicalphysics}. In the present case, the FT (\ref{LDFFT}) can be used to obtain non-trivial relations between the mechanical response of a physical system and its electrical and/or thermal transport properties. 

In absence of mechanical driving, the mechanical power vanishes, $\dot w =0$, and one recovers the usual steady-state FT for the currents \cite{Lebowitz_1999_JournalofStatisticalPhysics, Andrieux_2007_JournalofStatisticalPhysics}. On the other hand, a steady-state FT for the rate of mechanical work is recovered when considering a single heat reservoir \cite{Liu_2014_pre, Silaev_2014_PhysicalReviewE}.

Let us further mention the particular case of homogeneous temperatures, $\beta_\nu = \beta$, the FT (\ref{LDFFT}) then relates the fluctuations of mechanical power performed by the external driving to the chemical power performed by the particle currents, $\dot w_c \equiv \sum_\nu A^{n}_\nu j^{n}_\nu$. 

At the average level, the entropy production can be decomposed into
\be
\dot S_i = \dot S - \dot S_e \geq 0,
\ee
the positivity resulting from the FT for the entropy production. The average rate of system entropy change $\dot S$ is here given by the time derivative of the Shannon entropy in the Floquet basis, $S = \sum_s p_s \ln p_s$. The  average rates of entropy production and entropy flow are then given by \cite{Esposito_2012_PhysicalReviewE, Seifert_2012_ReportsonProgressinPhysics, VandenBroeck_2014_ArXive-prints}
\be
\dot S_i =   \sum_{ss'} \sum_{\nu l } \Gamma_{ ss'}^{\nu  l}  p_{s'}  \ln \frac{\Gamma_{ ss'}^{\nu  l} p_{s'}}{ \Gamma_{ s's}^{\nu \, -l} p_s }  
\ee
and
\be \label{entfl}
\dot S_e =  - \sum_{ss'} \sum_{\nu l } \Gamma_{ ss'}^{\nu  l}  p_{s'}   \ln \frac{\Gamma_{ ss'}^{\nu  l}}{ \Gamma_{ s's}^{\nu \, -l}}  
\ee
respectively. 

At this point, we emphasize the importance of distinguishing between microscopic processes involving different numbers of quanta exchanged with the external driving, especially when assessing thermodynamic properties. Indeed, a coarse-graining of the dynamics over the number of quanta exchanged with the external driving leads to a systematic underestimation of the entropy production \cite{Esposito_2012_PhysicalReviewE}. By using the log-sum inequality, one observes that
\be \label{coarsegrainedentropy}
\dot S^{cg}_i =   \sum_{ss'} \sum_{\nu  } \Gamma_{ ss'}^{\nu}  p_{s'}   \ln \frac{ \Gamma_{ ss'}^{\nu } p_{s'}}{ \Gamma_{ s's}^{\nu } p_s }   \leq \dot S_i ,
\ee
where the coarse-grained entropy production $\dot S^{cg}_i $ is written in terms of the coarse-grained transition rates $\Gamma_{ ss'}^{\nu  } =\sum_l \Gamma_{ ss'}^{\nu  l}$. We note that this coarse-graining can be understood as a coarse-graining of the dynamics in an extended Schnakenberg network whose micro-states correspond to individual Fourier modes of the Floquet states while the macro-states correspond to the Floquet states themselves \cite{Schnakenberg_1976_ReviewsofModernPhysics, Esposito_2012_PhysicalReviewE}.

At steady-state, the average rate of entropy change in the system vanishes, i.e. $\dot S = 0$, so that 
\be \label{rateofentropyproduction}
\dot S_i = -\dot S_e = -\sum_\nu \beta_\nu \dot Q_\nu = \beta_1 \dot W + \sum_{\nu} \left( A_{\nu}^{\epsilon} J_{\nu}^{\epsilon} + A_{\nu}^{n} J_{\nu}^{n} \right) ,
\ee
where we used the LDB condition (\ref{detailedbalance}) and the conservation laws for the currents and power at steady-state to obtain the last equality. This last expression shows that the average irreversible entropy production can be written as the sum of the powers dissipated by the currents against the thermodynamic affinities (\ref{affinities}) and the dissipated mechanical power. This picture proves useful at the time of characterizing the efficiency of thermodynamic engines as illustrated on the example exposed in the next section.

This finalizes the stochastic thermodynamic analysis of the model Hamiltonian introduced in section \ref{Hamiltonian}. The key points of the analysis are the following.

Though the current statistics is shown to be independent of quantum coherences in the Floquet basis, this is not the case for the mechanical work statistics at finite times. This is generally understood in the context of counting statistics by the fact that the quantum operator which is used to count mechanical work, $H_S (t) + \sum_\nu H_\nu$, does not commute in general with the density matrix of the system when the counting experiment begins. We note that this is in contrast to the slow driving situation, in which the environment naturally projects the system onto instantaneous eigenstates of the system Hamiltonian.

Nevertheless, the steady-state fluctuations of mechanical power are shown to be independent of quantum coherences in the Floquet basis. Within this limit, the contribution of the initial and final measurements becomes negligible, and the rate of dissipated mechanical work is equal to the rate of injection of quanta from the external driving to the system.

Despite the presence of coherences at finite times, we have shown that a thermodynamically consistent definition of entropy production can be introduced which only depends on the populations and their dynamics. As we explained, this peculiar property is mainly due to the dynamical decoupling between populations and coherences resulting from the use of the RWA.


\section{Model system} \label{models}


\begin{figure}[htbp]
\centerline{\includegraphics[width=10cm]{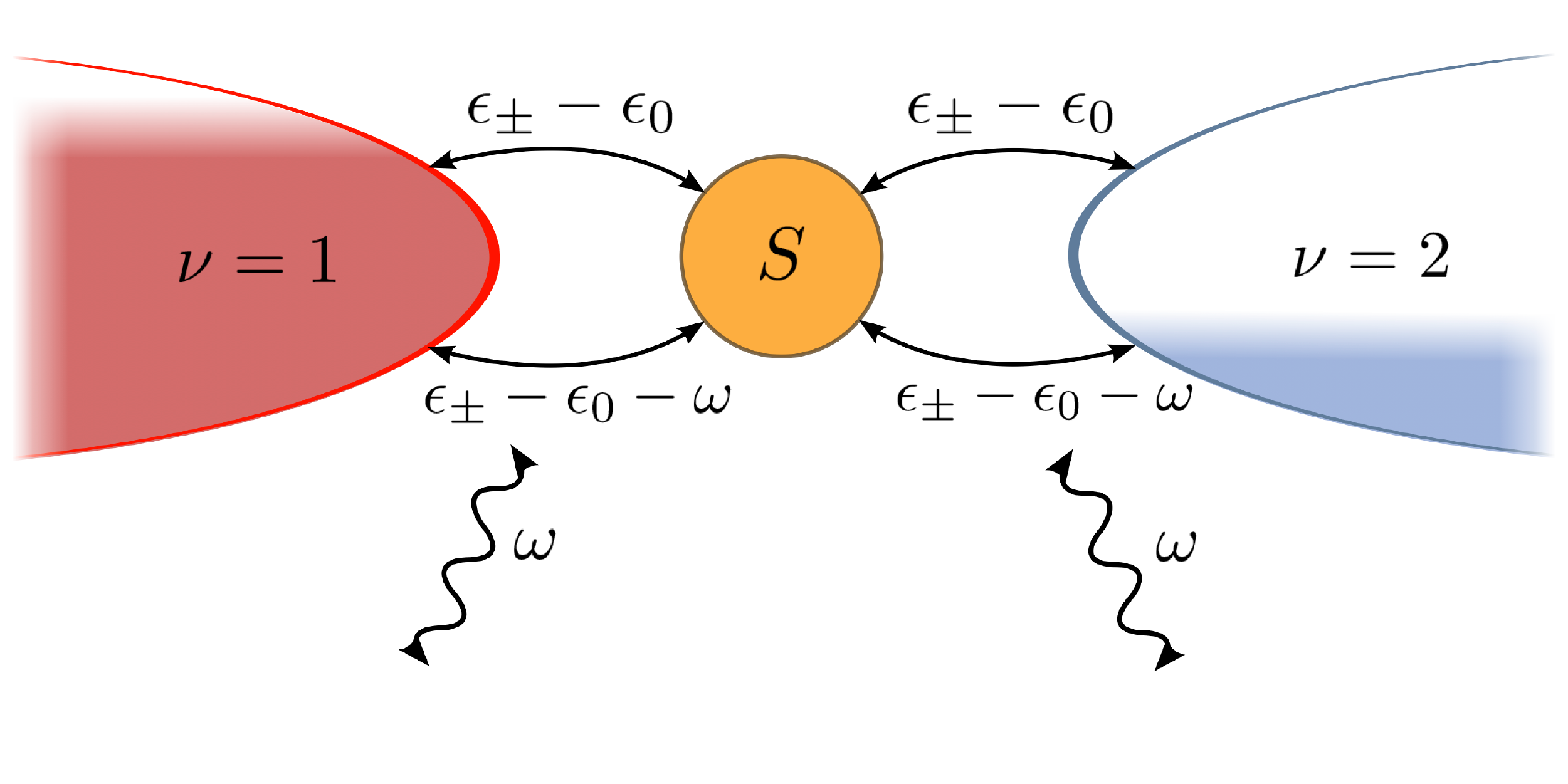}}
\caption{Schematic picture of the ac-driven QD connected to two particle reservoirs. The particle transfer processes with the reservoirs can be separated into two categories: those involving the absorption/emission of exactly one quantum of energy $\omega$ by the driving, and those that do not.}
\label{schematicpicture}
\end{figure}

We now make use of the analysis developed above in the study of a thermodynamic engine based on an ac-driven quantum dot (QD) coupled to two particle reservoirs \cite{Blumel_1991_PhysRevA, Cavaliere_2009_PhysRevLett, Wu_2010_PhysicalReviewB}. A schematic picture of the system is given in figure \ref{schematicpicture}. The ac-driven QD is conveniently modeled by the time-dependent Hamiltonian
\be
H_{\rm S} (t) = \frac{1}{2} \omega_0 ( | \uparrow \rangle \langle \uparrow | - | \downarrow \rangle \langle \downarrow | )+ \frac{\mu F}{2} \left( \mbox{e}^{- i \omega t} | \uparrow \rangle \langle \downarrow |
+ \mbox{e}^{i \omega t} | \downarrow \rangle \langle \uparrow | \right)
\ee
in terms of the splitting $\omega_0$ between the two single particle states of the system in absence of driving, $| \uparrow \rangle$ and $| \downarrow  \rangle$, the coupling strength $\mu F$ to the laser field and its frequency $\omega$.

The Floquet states of this system are readily obtained as 
\be
| \pm \rangle = \frac{1}{\sqrt{2\Omega}} \left( 
\pm \sqrt{\Omega \pm \delta}  | \uparrow \rangle + \sqrt{\Omega \mp \delta}\, {\mbox e}^{i \omega t}
| \downarrow \rangle
 \right)
\ee
in terms of the detuning parameter $\delta = \omega_0 - \omega$ and the Rabi frequency $\Omega = \sqrt{\delta^2 + (\mu F)^2}$. Their corresponding  quasi-energies are given by
\be
\epsilon_{\pm}  = \frac{\omega \pm \Omega}{2}.
\ee

The particle number in the QD fluctuates as a consequence of its interaction with the particle reservoirs. In the present case, the system is allowed to be in either the empty state $| 0 \rangle$ or the singly occupied states $|+\rangle $ and $|-\rangle$.

The interaction between the QD and the particle reservoirs is then modelled by the interaction Hamiltonian
\be
V = \sum_{\sigma= \uparrow, \downarrow} \sum_{\nu = 1,2 } \sum_k T^{\sigma}_{\nu k} ( c_{\nu k} |\sigma \rangle \langle 0 | + c_{\nu k}^\dagger  |0 \rangle \langle \sigma |  )
\ee
where $c_{\nu k}$ ($ c_{\nu k}^\dagger$) denotes the annihilation (creation) operator of a single particle state with wave number $k$ and energy $\epsilon_k$ in reservoir $\nu$, and $T^{\sigma}_{\nu k}$ is a parameter characterizing the strength of the coupling to the same reservoir. The reservoirs are themselves assumed to be composed of a collection of single particle states with Hamiltonian given by $H_{\nu} = \sum_k \epsilon_k c^{\dagger}_{\nu k} c_{\nu k}$.

The transition rates (\ref{transitionrates}) for this model can be evaluated by using the method described in section \ref{counting_statistics} yielding
\bea
\Gamma^{\nu, 0}_{0 \pm} & = & \frac{| \Omega \pm \delta | }{2 \Omega} \gamma^{\uparrow}_{\nu} (\epsilon_{\pm} - \epsilon_0) ( 1 - f_\nu (\epsilon_{\pm} - \epsilon_0)) \\
\Gamma^{\nu, 0}_{ \pm 0} & = & \frac{| \Omega \pm \delta |}{2 \Omega} \gamma^{\uparrow}_{\nu} (\epsilon_{\pm} - \epsilon_0)  f_\nu (\epsilon_{\pm} - \epsilon_0) \\
\Gamma^{\nu, -1}_{0 \pm} & = &\frac{| \Omega \mp \delta | }{2 \Omega} \gamma^{\downarrow}_{\nu} (\epsilon_{\pm} - \epsilon_0 - \omega) ( 1 - f_\nu (\epsilon_{\pm} - \epsilon_0 - \omega)) \\
\Gamma^{\nu, 1}_{ \pm 0} & = & \frac{| \Omega \mp \delta | }{2 \Omega} \gamma^{\downarrow}_{\nu} (\epsilon_{\pm} - \epsilon_0 - \omega) f_\nu (\epsilon_{\pm} - \epsilon_0 - \omega)
\eea
where the energy dependent tunneling rates are given by $\gamma^{\sigma}_\nu (x) \equiv 2 \pi  \sum_k |T^{\sigma}_{\nu k}|^{2} \delta (x - \epsilon_k)$. The Fermi-Dirac distributions $f_\nu (x) = (\exp{\beta ( x - \mu_\nu) } + 1)^{-1}$ characterize the statistical occupation of single particle states in reservoir $\nu$. As expected, these transition rates satisfy the LDB condition (\ref{detailedbalance}) 
\be
\ln \frac{\Gamma^{\nu, 0}_{ \pm 0}} {\Gamma^{\nu, 0}_{0 \pm}} =  \beta_\nu ( \epsilon_{\pm} - \epsilon_0 - \mu_\nu ) , \qquad \ln \frac{\Gamma^{\nu, 1}_{ \pm 0}} {\Gamma^{\nu, -1}_{0 \pm}} =  \beta_\nu ( \epsilon_{\pm} - \epsilon_0 - \omega - \mu_\nu )  .
\ee
We further note that the transitions with $l=1$ involve the exchange of smaller amounts of energy with the reservoirs as compared to those with $l=0$. This remark will have its importance when we later identify the best working regime of a thermodynamic engine based on this setup.

As a result of the non-equilibrium constraints applied to the system, the chemical bias $\Delta \mu = \mu_1 - \mu_2 $ and the periodic mechanical driving with frequency $\omega$, the system is subject to steady fluxes of energy and matter. These lead to a positive rate of entropy production (\ref{rateofentropyproduction}) here given by
\be \label{EPmodel}
\dot S_i = \beta \left(  \dot W + \dot W_c  \right) \geq 0.
\ee 
In this last equation, $\dot W$ denotes the average mechanical power provided by the ac-driving, while the quantity $\dot W_c = \Delta \mu \, J_{1}^n $ is the rate of chemical work provided by the current $J_{1}^n$ in order to bring particles from reservoir $1$ to reservoir $2$.

\begin{figure}[htbp]
\centerline{\includegraphics[width=10cm]{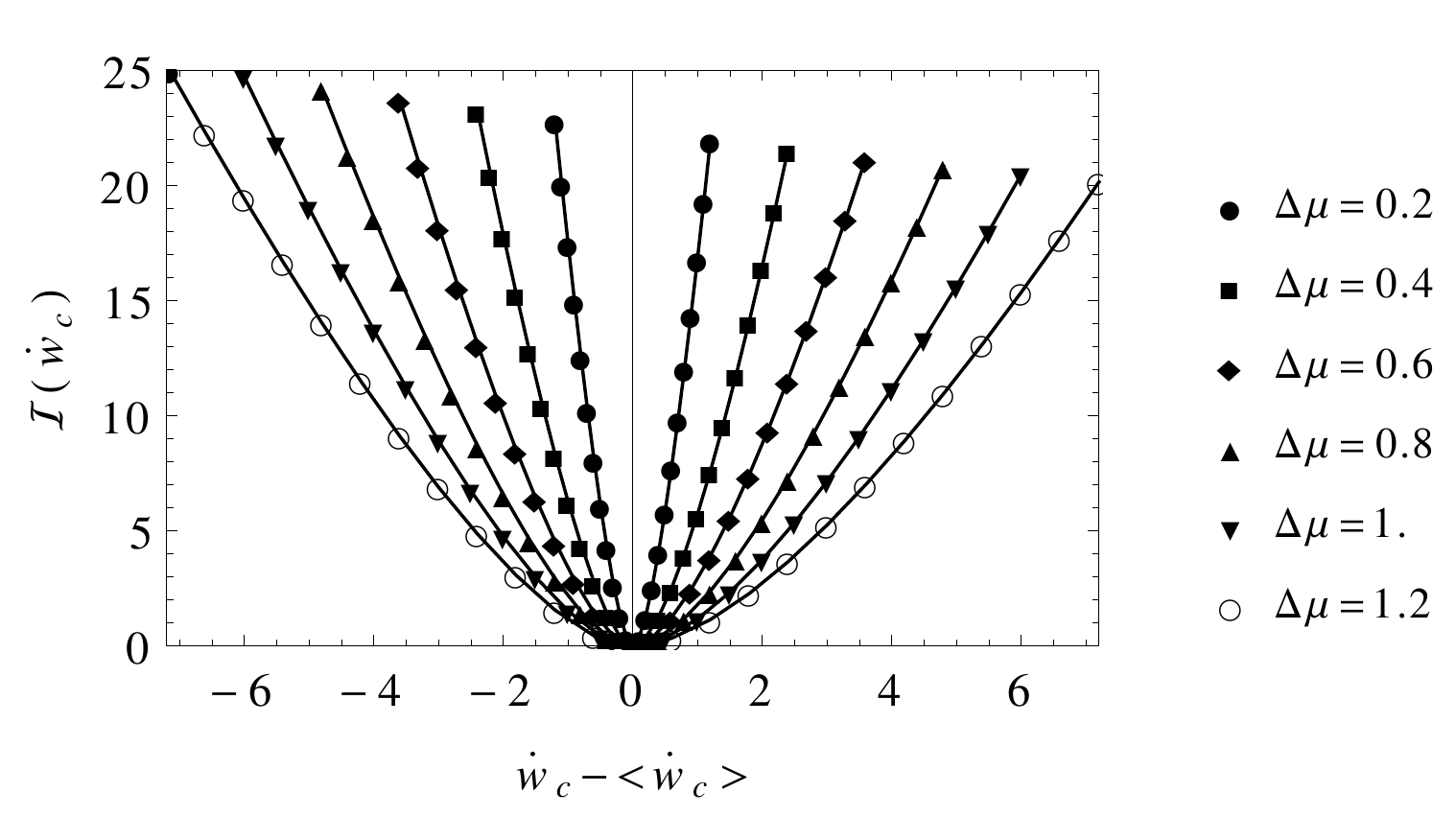}}
\caption{Illustration of the output power LDF for different values of the bias $\Delta \mu$ applied to the circuit. Other parameters where chosen as $\beta = 1$, $\mu \equiv (\mu_1 + \mu_2 ) /2 = 1 $, $\mu F = 1.4$, $\omega_0 = 0.7$, $\omega = 1.2$, $\gamma^{\sigma}_{\nu} \equiv \gamma = 1.3 $. Markers represent values obtained numerically, continuous lines being guides for the eye.}
\label{LDFfigure}
\end{figure}

The statistical properties of the dissipating fluxes $\dot w$ and $\dot w_c = \Delta \mu j_{1}^n$ are fully captured by their CGF or LDF as discussed in section \ref{counting_statistics}. Both were evaluated numerically and shown to satisfy the steady-state FTs (\ref{FTcurrentwork}) and (\ref{LDFFT}). For example, the joined LDF fo the mechanical and chemical powers $\mathcal{I} (\dot w , \dot w_c)$ was shown to satisfy the steady-state FT
\be
\mathcal{I} (\dot w , \dot w_c) - \mathcal{I} (-\dot w , -\dot w_c) = \beta (\dot w + \dot w_c),
\ee
in consistency with (\ref{FT2}). The right-hand side of this last relation is the fluctuating rate of entropy production of the system (cf. eq. (\ref{EPmodel})). In figure \ref{LDFfigure} we illustrate the marginal LDF of the chemical power, $\mathcal{I} (\dot w_c)$, for different values of the bias applied to the circuit.

We now consider a thermodynamic engine based on this setup which converts the input mechanical power $\dot w_{in} = \dot w$ performed by the external ac-driving into an output chemical power $\dot w_{out} = - \Delta \mu \, j_{1}^{n}$ provided to the particle current which now works against the chemical bias $ \Delta \mu >0$. The efficiency of such machine is defined as the ratio of its average output power divided by the average input power
\be \label{efficiency}
\eta = \frac{\dot W_{out}}{\dot W_{in}} = - \frac{\dot W_c}{\dot W} \leq 1,
\ee
the last inequality resulting from the second law of thermodynamics. 

The upper bound in (\ref{efficiency}) is only reached for vanishingly small output and input powers, i.e. close to equilibrium. This has motivated the investigation of the maximum output power regime, with regard to practical implementations \cite{Curzon_1975_AmericanJournalofPhysics, VandenBroeck_2005_PhysicalReviewLetters, Tu_2008_JournalofPhysicsA, Esposito_2009_Physicalreviewletters, Seifert_2011_PhysicalReviewLetters}. Two  aspects must be considered to attain this regime. 

One is the identification of the properties of the external system to which our engine will provide the highest output power. In the present case, the external system consists of the circuit formed by the reservoirs themselves, and its adjustable parameter is the output bias $  \Delta \mu$.

The design of the system performing the conversion and its connection to the environment constitute other important aspects of power optimization. Here, the quantum dot itself is the vector of the conversion and its spectrum and interaction parameters with the driving and reservoirs provide the adjustable parameters in order to reach maximum output power. In particular, we note that an asymmetry in the coupling between the system and reservoirs $1$ and $2$ is necessary for the conversion from mechanical to chemical work to be possible. An extreme and ideal situation is the one for which the input and output powers are tightly coupled, i.e. $\dot w_{out} \propto \dot w_{in}$, and are thus maximally correlated. 

In the following, we consider our engine to work in the tight coupling regime with the only non-vanishing tunnelling amplitudes being $\gamma_{1} (\epsilon_\pm - \epsilon_0 )$ and $\gamma_{2} (\epsilon_\pm - \epsilon_0 - \omega )$. In this situation, the mechanical driving provides one quantum of energy equal to $\omega$ to charge the system from reservoir $ 2$, while the system can be discharged into reservoir $ 1$ without any energy supply from the environment. This favors the net pumping of particles from reservoir $2$ to reservoir $1$ against the bias $\Delta \mu >0$. 

In this regime, the output and input powers are proportional to each other so that the engine efficiency can be simply written as
\be
\eta = \frac{\Delta \mu}{\omega}.
\ee

Within the regime of maximal output power, most studies have focused on the average output power and the corresponding efficiency of the considered engine. Here, we use the counting statistics formalism exposed in section \ref{counting_statistics} above in order to investigate the fluctuations of output power. 

\begin{figure}[htbp]
\centerline{\includegraphics[width=15cm]{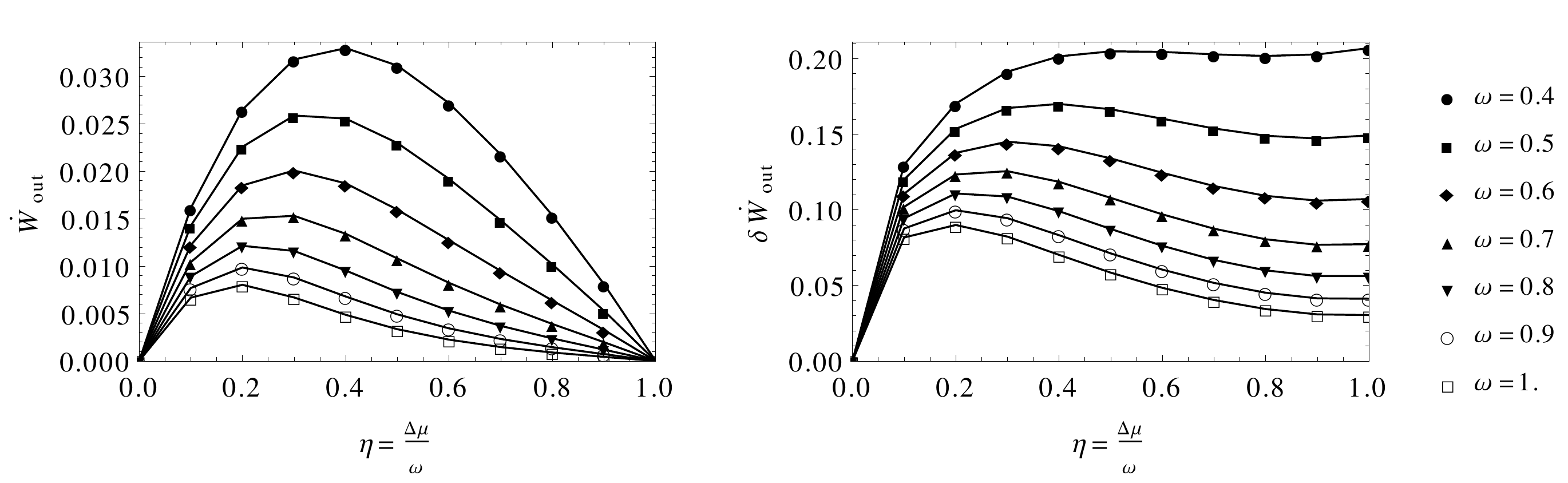}}
\caption{Average (left panel) and root mean square (right panel) of the output chemical power $\dot w = - \dot w_c$ performed against the bias $\Delta \mu >0$ as a function of the engine efficiency $\eta = \Delta \mu /\omega$ and for different values of the driving frequency $\omega$. Parameters are chosen as $\beta = 1$, $\mu \equiv (\mu_1 + \mu_2)/2 = 1$, $\mu F = 2$, $\omega_0 = 0.7$, $\gamma_{1} (\epsilon_\pm - \epsilon_0 ) = \gamma_{2} (\epsilon_\pm - \epsilon_0 - \omega ) = 1.3 $.}
\label{outputmeanandfluct}
\end{figure}

The average and mean root square of the output power, respectively given by 
\be 
\dot W_{out} = -\Delta \mu \, J_{1}^{n} \quad \mbox{and} \quad \delta \dot W_{out} =\Delta \mu \sqrt{\langle (  j_{1}^{n} -   J_{1}^{n}) ^{2} \rangle } 
\ee
are illustrated in figure \ref{outputmeanandfluct} as a function of the efficiency (\ref{efficiency}). The regimes of maximum output power are shown to correspond to high, although not necessarily maximal, power fluctuations. Another striking feature is the relative size of power fluctuations which are one order of magnitude larger than the average power in the illustrated regime.

\begin{figure}[htbp]
\centerline{\includegraphics[width=9cm]{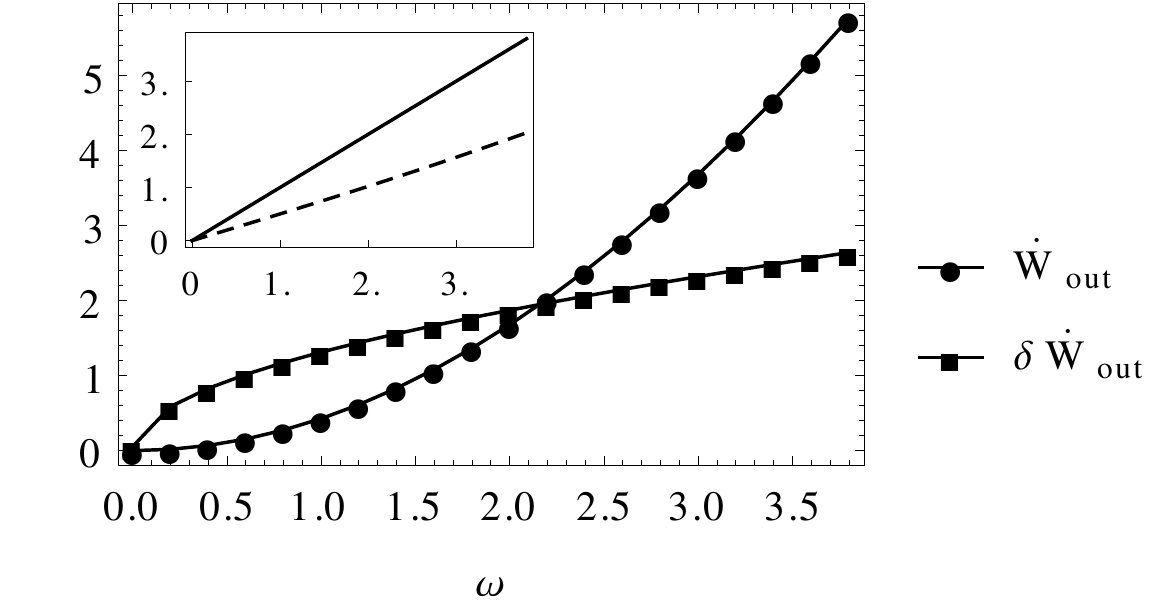}}
\caption{Plots of the average and root mean square of output power in the regime of maximum output power. Parameters are chosen as $\beta = 1$, $\gamma_{1} (\epsilon_\pm - \epsilon_0 ) = \gamma_{2} (\epsilon_\pm - \epsilon_0 - \omega ) = 20 $. Other parameters are numerically adjusted to reach maximum output power. $\omega_c$ is here defined as the frequency value at the crossing point between the average current and its variance. Inset: (solid line) optimized values of the natural frequency of the QD $\omega_0$ as a function of the driving frequency $\omega$, (dashed line) optimized values of the ouput bias $\Delta \mu$ as a function of the driving frequency.}
\label{schematicpicture}
\end{figure}

In figure \ref{schematicpicture}, parameters of the ac-driven QD and the output bias $\Delta \mu$ were individually adjusted to reach maximum output power for each value of the driving frequency $\omega$, while temperature was fixed at $\beta = 1$. The magnitude of power fluctuations are seen to remain significantly bigger than the average up to a certain value of the driving frequency. Above this value, the situation is reversed and the average and root mean square output power increase, respectively, quadratically and linearly with the driving frequency. The optimize values of the bias $\Delta \mu$ and the natural frequency $\omega_0$ of the QD increase linearly with the driving frequency $\omega$, as shown in the inset of the same figure. The regime of large driving frequency can thus be here understood as a low temperature limit in which fluctuations are indeed expected to be reduced. This is confirmed by repeating this maximization procedure for different values of temperature. By doing this, one sees that the value of the driving frequency $\omega_c$ at the crossing point, where the average value of the output power and its variance are equal, increases monotonically with temperature.

The above observations may be summarized as follows. First, the fluctuations of output power in the quantum engine we consider can be substantial and exceed its average by more than one order of magnitude. This assertion is in the present case particularly justified in the regime of maximum output power for which the fluctuations are shown to systematically exceed the average output power up to a certain value of the driving frequency. Second, the relative magnitude of fluctuations with respect to the average may be lowered by getting away from equilibrium.

These observations suggest that compromises may be necessary in the design of nano-scaled heat engines, depending on the priority to deliver a high or stable output power. In this sense, engines perform better far rather than close to equilibrium.


\section{Conclusion and perspectives} \label{conclusion}


We reported the stochastic thermodynamics analysis of a weakly coupled open quantum system connected to multiple reservoirs and driven by a fast external field. This analysis is of particular interest in the context of the study of nano-scaled thermodynamic engines as illustrated in the example we considered in the previous section. Such models of thermodynamic engines may be realized through ac-driven semiconducting circuits \cite{Blumel_1991_PhysRevA, Cavaliere_2009_PhysRevLett, Wu_2010_PhysicalReviewB} or cold atom gases \cite{Krinner_2013_PhysRevLett, Brantut_2012_Science, Brantut_2013_Science, Gallego-Marcos_2014_ArXive-prints}.

The use of the rotating wave approximation (RWA) has proven useful in order to gain physical insight into the thermodynamic properties of open quantum systems as was already the case for the slow driving limit. However, several open issues remain regarding the thermodynamic properties of driven open quantum systems outside the range of application of the RWA.

In particular, the characterization of the entropy production in quantum systems with sustained coherences and their subsequent thermodynamic analysis constitute challenging problems of non-equilibrium quantum thermodynamics.

Another issue is the identification of thermodynamic quantities when broadening effects cannot be neglected. On top of these fundamental difficulties, the expansion to second order in the interaction between system and reservoirs is void in this case, and one must appeal to complementary methods such as the non-equilibrium Green's functions formalism \cite{Esposito_2015_PhysicalReviewLetters}.


\section{Acknowledgment} \label{acknowledgment}


The authors would like to thank Martin Holthaus from the Carl von Ossietzky Universit\"at Oldenburg for stimulating discussions.

The research was supported by the National Research Fund, Luxembourg in the frame of project FNR/A11/02 and of the AFR Postdoc Grant 7982468.



\section*{References}


\label{Bibliography}


\bibliography{physics}


\section{Appendix} \label{appendix}


We here prove that the mechanical work CGF is independent of quantum coherences in the Floquet basis for systems with bounded energy.

We first note that $\rho_{ss'}(\xi_\nu , \lambda_\nu ,t) \rightarrow 0$ for $t\rightarrow \infty$ by virtue of (\ref{coherencesevol}). By using this property and expressions (\ref{transientworkGF}) and (\ref{workCGF}), we rewrite the work CGF as
\be
\mathcal{G}_w (\alpha) = \lim_{t\rightarrow \infty} \frac{1}{t} \ln \sum_{e_0 e_t} \mbox{e}^{-\alpha \Delta \epsilon_S} \sum_s  \langle e_t | s \rangle   \rho_{ss} (i\alpha, t; e_0) p_{e_0}  \langle s| e_t \rangle .
\ee

The populations $ \rho_{ss} (i\alpha, t; e_0) $ satisfy a master equation similar to (\ref{modmasteq}) but with rate matrix given by (\ref{wordrm}). A formal solution of this equation can be written as 
\be
\boldsymbol{\rho} (i\alpha, t; e_0)  = \mbox{e}^{\boldsymbol{\Gamma } (\alpha) t} \cdot \boldsymbol{\rho} (0 ; e_0)
\ee
where the initial condition is chosen as $\rho_{ss} (0 ; e_0) = \langle s | e_0 \rangle \langle e_0 | s  \rangle $, $ \rho_S (0)$ denoting the reduced density matrix of the system before the initial measurement of the system energy takes place.

We now introduce the eigenvectors $\boldsymbol{v} (\alpha)$ of the rate matrix $\boldsymbol{\Gamma } (\alpha) $ with eigenvalues $v (\alpha)$ such that $\boldsymbol{\Gamma } (\alpha)  \cdot \boldsymbol{v} (\alpha) = v (\alpha)  \boldsymbol{v} (\alpha)$. The CGF can now be rewritten as
\be \fl
\mathcal{G}_w (\alpha) = \lim_{t\rightarrow \infty} \frac{1}{t} \ln\sum_{e_0 e_t} \mbox{e}^{-\alpha \Delta \epsilon_S} \sum_v   \sum_{s \tilde s }  \langle e_t | s \rangle   \langle s| e_t \rangle v_s (\alpha) v_{\tilde s}^{*} (\alpha) \mbox{e}^{v (\alpha ) t}   \rho_{\tilde s \tilde s} (0; e_0)
\ee
where the sum over $v$ runs over all the eigenvectors of $\boldsymbol{\Gamma } (\alpha)$.

By assuming the system to have a bounded energy at all times, $\Delta \epsilon_S$ becomes negligible in front of $v (\alpha) t$ as $t \rightarrow \infty$, and we get that
\bea
\mathcal{G}_w (\alpha)&  = &  \lim_{t\rightarrow \infty} \frac{1}{t} \ln\sum_{e_0 e_t} \sum_v   \sum_{s \tilde s }  \langle e_t | s \rangle   \langle s| e_t \rangle v_s (\alpha) v_{\tilde s}^{*} (\alpha) \mbox{e}^{v (\alpha )t}   \rho_{\tilde s \tilde s} (0; e_0) p_{e_0} \\
 & = &   \lim_{t\rightarrow \infty} \frac{1}{t} \ln\sum_{e_0 } \sum_v   \sum_{s \tilde s }   v_s (\alpha) v_{\tilde s}^{*} (\alpha) \mbox{e}^{v (\alpha )t}   \rho_{\tilde s \tilde s} (0; e_0) \\
 & = & \overline{v} (  \alpha)
\eea
where $ \overline{v} (  \alpha)$ is the dominant eigenvalue of the rate matrix $\boldsymbol{\Gamma } (\alpha)$, i.e. $ \overline{v} (  \alpha) = \mbox{max}_{v(\alpha)}  \left\{ \mbox{Re} \left\{ v (\alpha) \right\} \right\}$.

\end{document}